# Misalignment or misuse? The AGI alignment tradeoff

Max Hellrigel-Holderbaum and Leonard Dung[1]

Abstract

Creating systems that are aligned with our goals is seen as a leading approach to create safe and beneficial AI in both leading AI companies and the academic field of AI safety. We defend the view that misaligned AGI – future, generally intelligent (robotic) AI agents – poses catastrophic risks. At the same time, we support the view that aligned AGI creates a substantial risk of catastrophic misuse by humans. While both risks are severe and stand in tension with one another, we show that – in principle – there is room for alignment approaches which do not increase misuse risk. We then investigate how the tradeoff between misalignment and misuse looks empirically for different technical approaches to AI alignment. Here, we argue that many current alignment techniques and foreseeable improvements thereof plausibly increase risks of catastrophic misuse. Since the impacts of AI depend on the social context, we close by discussing important social factors and suggest that to reduce the risk of a misuse catastrophe due to aligned AGI, techniques such as robustness, AI control methods and especially good governance seem essential.

## 1. Introduction

Take AGI (artificial general intelligence) to be hypothetical future AI which is in all or virtually all cognitive capacities (reasoning, remembering, planning etc.) superior to humans, such that it can perform every kind of intellectual work (scientific research, writing, stock market trading, etc.) much better than humans.[2] Here, we assume that there is a realistic chance that AGI will be created soon enough to necessitate serious thinking about its consequences (for some relevant views, see Barnett 2023; Cotra 2020; Davidson 2021a, 2021b; Grace et al. 2023; Roodman 2020).

Our subsequent arguments presuppose that AGI pursues goals, in some sense. We take it that we can speak of goals even in current AI systems (see Dung 2023), but if one denies this, one can simply stipulate that AGI – in the sense relevant here – is a goal-directed autonomous agent and (in our view plausibly) point out that there are strong incentives favoring the creation of AI agents, rather than mere AI tools (Chan et al. 2023). Finally, there are many tasks that AI systems could best solve by physically interacting with the environment, thus creating incentives for developing AGI robots. The arguments we develop here are not limited to robots; however, many of the risks we raise are particularly salient and pressing for AGI robots.[3]

---

[1] Both authors contributed equally to this paper.

[2] Such systems may also be called "superintelligence" (Bostrom 2014). Moreover, many of our considerations concern superintelligence in a strict sense, as systems whose cognitive capacities vastly exceed all humans.

[3] Discussions of misalignment in the AI safety literature also reflect the saliency of robots: They usually concern agents which are delegated to act in the physical world on behalf of a human (Amodei et al. 2016; Hadfield-Menell et al. 2016; Kenton et al. 2021; Russel 2019). Our arguments likewise presuppose that AGI systems can, at least indirectly, affect the physical world. While we expect AGI robots to be the ultimate solution for this, in

If AGI is created, it is highly important that it is designed in a safe manner. The main contribution of leading AI companies to help achieve this is AI alignment research. As a first approximation, an AI system is aligned, in a narrow technical sense, if it tries to do what its designers/users want it to do (Christiano 2018; Dung 2023; Ngo et al. 2024).[4] AI that always pursues human goals is intended to be safe.

In this paper, we analyze and discuss what we call the *AGI alignment dilemma*. That there is an AGI alignment dilemma is suggested by the following contributions: A rich body of literature has developed arguments suggesting that misaligned AGI will likely exhibit instrumentally convergent behavior and since power-seeking is among those, ultimately try to disempower humanity (Bostrom 2012, 2014; Omohundro 2008). According to some authors, this could cause catastrophic outcomes, perhaps even human extinction (Carlsmith 2022; Cotra 2022; Dung 2024a; Langosco et al. 2022; Ngo et al. 2024). However, other authors, such as Friederich (2023) and Yum (2024), have argued forcefully that aligned AGI creates severe risks of misuse through humans, which will likely cause catastrophic outcomes. Thus, there is some reason to think that a lack of alignment entails catastrophic risks of human disempowerment from AGI while success at alignment implies catastrophic risks from AGI misuse by humans. The dilemma consists in the decision between aiming to create aligned AGI or allowing for AGI misalignment, given that both threaten catastrophic outcomes.

Section 2 motivates this dilemma in more detail. In section 3, we discuss objections to the AGI alignment dilemma. Such objections deny part of the dilemma, holding that one of the suggested horns is not actually problematic. While we concede plausible scenarios in which these risks do not materialize, we argue that the risks remain substantial. Section 4 starts investigating the relation between alignment and misuse risks empirically based on current techniques and misuse scenarios, leading us to conclude that they trade off as some influential alignment techniques plausibly facilitate misuse. Section 5 discusses the question whether social factors may improve catastrophic takeover risk without increasing catastrophic misuse risk and vice versa as well as broader social factors which substantially shape or constitute the AGI risk landscape. Finally, section 6 concludes. Although this paper focuses on the alignment and misuse of AGI, the same tradeoff might exist for the alignment of current AI systems, albeit with a reduced severity of both risks.

---

the meantime AGI would likely use humans or narrow, remote-controlled robots (e.g. drones) to this effect, raising similar challenges.

[4] The technical question *how* to align AI with certain goals and values can be distinguished from the ethical question about the goals or values AI *should* be aligned with (Gabriel 2020).

## 2. The AGI alignment dilemma

We now consider both horns of the AGI alignment dilemma in more detail. Suppose that an AGI is misaligned. This means that its goals do not correspond to what its designers want them to be, making it a superhuman system which optimizes for aims which are different from what its designers intended. It seems plausible to us, and many others, that a sufficiently intelligent system would be more powerful than humanity and able to dominate it, if it wanted to (Bostrom 2014; Russell 2019). This is especially plausible for superintelligent robots. Let us assume this AGI has the capacity to dominate humanity. Researchers have argued that there is a high risk that such a misaligned system may disempower humanity (Carlsmith 2022; Cotra 2022; Dung 2024a).

The most common reason for believing this is encapsulated in the instrumental convergence thesis (henceforth ICT)[5], most influentially supported by Bostrom (2012, 2014, p. 131 ff.) and Omohundro (2007, 2008).[6] According to the ICT, there are certain goals which are highly instrumentally useful for a wide range of final goals. The accumulation of power and resources is taken to be one such convergent instrumental goal since power and resources are generally useful for achieving a wide range of other goals. Bostrom and others reason that AGI, for a wide range of goals, would have an incentive to accumulate power and resources as those generally raise the probability that its final goals will be satisfied which is why we should expect it to engage in such behaviors.[7] Following this line of thinking: If AGI dominates humanity, then it can achieve whatever other final goal it happens to have, no matter whether humans like it, and it can do so without human interference. Since this seems to provide a good instrumental reason to disempower humanity and since AGI, by our stipulation, is capable of disempowering humanity, it will do so. Researchers tend to think that such an AI takeover would be an ethical catastrophe (Carlsmith 2022; Dung 2024a). This is particularly plausible if AI takeover leads to a loss of humanity's control over its future, a disenfranchisement of all humans depriving us of the capacity to lead flourishing lives, or human extinction (Bales 2024). Call this an *AGI takeover catastrophe*, the first horn of the dilemma.

---

[5] Note that this has nothing to do with Information and Communications Technology, also commonly abbreviated "ICT".

[6] While this is the most prominent and widely supported reason for AGI posing catastrophic risk (due to misalignment), several others have been raised. Zhuang and Hadfield-Menell (2021) argue for catastrophic risks from misalignment without relying on instrumental convergence. Bucknall and Dori-Hacohen (2022) focus on how current and near-term AI may contribute to existential risks. Hendrycks et al. (2023), Vold and Harris (2021) and Critch and Russell (2023) provide overviews that include several other potentially catastrophic risks.

[7] Note that on the most straightforward interpretation, there are several empirical cases of relevant types of instrumentally convergent behavior in state of the art AI systems. Examples include (attempted) deactivation of oversight/control mechanisms, self-preservation, scheming, and strategically acting as though being in agreement with the training objective (Greenblatt et al. 2024; Lu et al. 2024; Meinke et al. 2024; OpenAI 2024; Scheurer et al. 2024). One important source of uncertainty for interpreting these cases concerns how accurately models' chains-of-thought represent their internal computations.

The typical response proposed in the literature, and worked on by leading AI labs, is AI alignment (Aschenbrenner 2023; Ji, Qiu, et al. 2024; Leike et al. 2022). If the final goals of AGI correspond, in some sense, to human goals, then – so the thought goes – AGI will not disempower humanity since this would conflict with its final goals.[8] For instance, aligned AGI might have the final goal to save humans from harm, on some reasonable definition of harm. If it robustly pursues this goal, it will not make humanity go extinct, because this would harm humans.

However, aligned AGI poses catastrophic risks as well (Friederich 2023; Yum 2024). A fully aligned AGI system, by definition, pursues human goals. Yet, this means that the AGI system will optimize for bad outcomes if the relevant humans have thoughtless or even malicious goals. Since, by our stipulation, AGI is powerful enough to dominate all of humanity, humans "with a monopoly, or near-monopoly, of aligned, or near-aligned, AGI, may well have power that is far superior to that provided by any technology today" (Friederich 2023, sect. 4). Having some humans control AGI would lead to an unparalleled form of power concentration. These humans could take control of humanity if they so choose, and they could do it in whatever way they prefer. For instance, they could use AGI to covertly take control, or to manipulate populations into willingly accepting them as leaders. Plausibly, no human can be trusted with that amount of power. Moreover, if multiple humans have access to an AGI, any one of them could try to use it to subjugate the rest of humanity. Call such outcomes an *AGI misuse catastrophe*, the second horn of the dilemma.

Crucially, it pre-theoretically seems like an AGI misuse catastrophe depends on having sufficiently aligned AGI. If AGI is not aligned, at least to a certain necessary degree which depends on the relevant context, then operators cannot control it and thus cannot bring it to reliably and systematically act in accordance with their goals. Thus, *intentional* plans to use AGI for their own purposes cannot be implemented. Operators cannot, for instance, use AGI to establish a totalitarian dictatorship, or maximize their power in other ways if the AGI is not aligned.

So, we have a candidate dilemma. If we build misaligned AGI, this threatens an AGI takeover catastrophe. If we build aligned AGI, this might cause an AGI misuse catastrophe. Both concerns do not rely on, but are especially pronounced with, robot AGI. In the rest of this paper, we examine this candidate dilemma. First, we discuss whether this dilemma is real. According to some objections, presented in the next section, one or both of these conditionals are false so that at least one of the suggested horns can be rejected. We develop novel responses to these objections. Moreover, also in section 3, we contend that the relation between catastrophic takeover and misuse risk should be understood as a context-dependent tradeoff that requires empirical investigation, rather than as an inescapable, conceptually provable dilemma.

---

[8] This is a broad sketch of the rationale for AI alignment which will be fleshed out in section 4 with more attention to the empirical alignment literature.

## 3. Is there a dilemma? Some objections

We think that the objections have some merit. It is – if we get lucky – possible to avoid both kinds of catastrophic outcomes from AGI. However, it is doubtful that they have sufficient force, so that the risk of a catastrophic outcome remains unacceptably high on both horns of the AGI alignment dilemma.

### 3.1 Misaligned AGI and risks of a takeover catastrophe

Let us begin with the first horn: If AGI arises and is misaligned, then this makes an AI takeover catastrophe likely. If one rejects this conditional, one needs to say that misaligned AGI is unlikely to engage in excessive power-seeking and thus to try to disempower humanity. The argument for the first horn was that disempowering humanity would be instrumentally very useful for AGI with a wide range of misaligned final goals, so that AGI would have a strong incentive to disempower humanity.

We think that the argument based on the ICT to the claim that misaligned AGI would lead to AGI takeover should at least be taken seriously, if not given a considerable credence. To show why, we will critically examine the two objections to this argument which seem strongest to us.[9] The following discussion is necessarily somewhat speculative: this is due to both the nature of the subject matter as well as the paucity of peer-reviewed literature on the topic. Moreover, while we discuss the objections that seem most plausible to us, there might be others, perhaps not yet developed, which undermine the argument based on the ICT. So, we merely claim that this argument is credible given the current state of research, not that it is established as sound yet. In our view, this would ultimately require extensive empirical evidence. Finally, if one is skeptical of the view that misaligned AGI risks a takeover catastrophe, one can take our arguments as conditional: They explore the consequences of this common (see the references above) view.

---

[9] Gallow (2024) gives a prominent objection that we do not address separately here despite its valuable contribution. This is because Gallow's paper mostly addresses non-sequential decision-making while we expect realistic cases of dangerous instrumentally convergent behaviors to concern cases of sequential decision making. The brief discussion of sequential decision making in Gallow's paper fundamentally does not contradict the ICT.

*Objection 1, against usefulness:* One may take issue with the assumption that power and other instrumental goals are sufficiently useful to incentivize catastrophic forms of power-seeking. Sharadin (2024) makes this case in an argument that can be summarized in the following two steps: 1) the best-supported notion of goal-promotion contrasts different kinds of behavior and their influence on the likelihood of achieving a goal instead of speaking of behaviors/incentives being useful tout court. 2) with such a notion, he argues that acting to accumulate extreme amounts of power/resources does not (generally) promote the relevant goal over moderate variants of such behavior. Hence, AI systems which optimize outcomes, relative to their final goals, will not be motivated to accumulate extreme amounts of, say, power (ibid.). In support of 2) one may add that an AGI may figure out more effective behaviors for achieving its final goal than general, indirect means like seeking power. There are generally benefits to specialization, including for AI systems (Drexler 2019; Shah 2023). Thus, the marginal benefits of specialized behavior may typically trump general strategies like seeking power. Additionally, while acting to disempower humanity has expected benefits with respect to most final goals, it also has expected costs (Sharadin 2024, sect. 5). If the costs outweigh the benefits, then AGI should not be expected to try to disempower humanity.

In response, one may note that, as Dung (2024a) has argued, informally, it seems that the benefits of disempowering humanity would be very high, while – for a sufficiently superior AGI – it should be easy and thus not costly. Plausibly, similar considerations hold for other candidate instrumentally convergent behaviors. Accumulating extreme amounts of resources plausibly maintains its usefulness (particularly for AGI that need not be confined to one body) as it enables a higher ability to manipulate physical or social reality and to process information. Reducing the chance (even slightly) that one's goals are changed or oneself is destroyed seems very useful for lots of goals as it maintains one's ability to act towards one's goal while it does not need to be exceedingly costly.

Further, it seems doubtful to us that there are no sufficient comparative benefits to massive extents of, say, power-seeking over moderate amounts. It has been argued that open-ended goals will plausibly develop in AGI due to strong incentives to develop AIs along such lines (Carlsmith 2022; A. Chan et al. 2023). The content of open-ended goals, for example *maximize pleasure*, contains no restrictions in time and space, in contrast to closed-ended goals like, e.g., *have fun tonight* (Hellrigel-Holderbaum 2024). It seems that extreme amounts of resources or power would indeed promote many open-ended goals over moderate amounts (see also Bales 2024).

Lastly, assume that there are more effective (specialized) means for pursuing a given goal than seeking power. This may nonetheless fail to reduce the chance much that seeking power as a less effective means is also pursued, if the latter still increases the chance that (or extent to which) the goal is achieved. Using an example from humans: The fact that drinking water may be most important for staying alive provides little evidence that humans will not also take care of their health and thereby their survival via less effective means. Even if there is a more effective means to its goal than seeking power, AGI could still seek power if it possesses a sufficiently large action repertoire per unit of time. If so, it may simply manage to pursue both the most effective means *and* seek power as a means of secondary importance to achieve its goal. Further, the most effective behavioral sequences possibly combine what we would consider a direct strategy towards a goal with power-seeking. In practice, action spaces are surely much more fine-grained than the rough categories we use to describe them so that AGI may find behaviors that are both directly useful and over time result in the accumulation of extreme amounts of power.

*Objection 2, against AGI as rational, coherent goal maximizer:* It may be that AGI will not, or not only, maximize expected value with respect to some well-defined final goal, but that the AGI's behavior will be determined by the interaction of a variety of competing, contextually activated heuristics which emerge during training (L. Chan 2022; Lang 2023).[10] One inspiration for this view is the hypothesis that model-free reinforcement learning (RL) leads to systems of this kind. In model-free RL, the RL system learns associations between states and actions based on previous experience (its policy). This can be understood as the RL system learning heuristics about which actions in which states yield a high reward in its environment. The reward is used to update the parameters of the model and thus the policy – without however constituting a goal of the system in the way that classical rational agents' goal is to maximize their expected utility (Turner 2022). It may then be that the model-free RL system's behavior should be understood as being fully determined by the competition of these heuristics, while the system does not maximize some consistent utility function. Arguably, aspects of human decision-making (roughly "system-1 reasoning") can be modelled in these ways and are consequently not well captured by a select number of final goals. If an AGI has this structure, then it may not indefinitely optimize for some final goal or instrumentally convergent goal like power, since goals are contextually constrained and embedded in a set of other competing behavioral tendencies.

---

[10] See also (Bales 2023) for some motivation for this objection. He examines why AGI may not conform to the classical axioms which imply the maximization of expected utility.

Our reply holds most generally that the deviations from AGI as a rational, coherent goal maximizer posited above do not imply an insensitivity to instrumental incentives and thus do not entail an absence of behaviors like power-seeking.

More specifically, we should expect AGI to be able to plan. There are several reasons for this. First, model-free RL does not exclude planning. The above characterization of model-free RL describes how the behavior and policy are adapted over time, but it does not describe the processes used to achieve that behavior. Hence, it is compatible with planning as planning can be internalized into the neural network used to estimate the policy or value function. Second, planning is a particularly flexible method for solving tasks, particularly compared to context-dependent behavioral heuristics (see e.g. Dolan and Dayan 2013; Niv et al. 2006). Since planning is both very general and useful, we should anticipate systems deserving to be called AGIs to be capable of it (Carlsmith 2022; A. Chan et al. 2023).[11] With planning abilities, AGI should recognize the value of instrumental behaviors like seeking power so that we are back to the initial argument.

Suppose one holds, despite this, that AGI may be developed that cannot plan, avoiding even internalized planning (defining AGI without referring to planning capacities). The relevant AGI would be guided by context-sensitive heuristics. Nonetheless, this AGI may still engage in dangerous forms of power-seeking. Again, deviations from instrumental rationality do not imply wholesale instrumental irrationality. Insofar as behaviors due to heuristics are sensitive to their usefulness to goals — and that is how they are selected in training — they may well include dangerous instrumental behaviors. If so, heuristics-bound instrumental rationality may be sufficient for excessive power seeking, particularly as formally speaking, seeking power is optimal for most reward functions (Turner et al. 2019). A case in point may be that plausibly, many humans would take control over humanity if they had the power to securely do so (Carlsmith 2022). This could, depending on the person, be due to learned heuristics in which having power over others is conducive to personal goals.[12]

Since both objections deny that a misaligned AGI must pursue instrumentally convergent goals, it may be worth giving a general remark about misaligned AI irrespective of the ICT. In general, having an autonomous, highly powerful system around, whose goals do not reflect humanity's wants and needs, – such as misaligned AGI – seems eminently dangerous (even more so if it can directly act on the physical world). A central reason is that since humans require very specific conditions to flourish and only very specific changes in circumstances are good for us, most possible goals of an AGI, and especially ones that are not aligned, would be bad news for humanity. As an analogy, suppose we would learn that superintelligent extraterrestrials will arrive at earth, and that they have the power to enslave or annihilate humanity, if they so choose. Suppose further that all theoretical arguments to predict the goals or behaviors of these aliens are flawed. By stipulation, they may or may not come in peace. This situation certainly seems much too dangerous for comfort. Creating misaligned AGI, without a deeper understanding of what it will be like, is thus highly undesirable even if the ICT does not hold.

---

[11] Indeed, it is common to include planning capacities in the definition of AGI, as we also did above.

[12] Alternatively, some humans may make plans to take control over humanity. If so, that may support the prior response: We should expect AGI to plan so that it picks up on incentivized behaviors like power-seeking.

There is nonetheless a silver lining to this discussion. We have seen some reasons to think that the amount of risk from misaligned AGI systems depends on their broad cognitive functioning: for example, whether they rely on a variety of context-dependent behavioral heuristics as opposed to functioning as general planning systems and whether their goals are open-ended or closed-ended. Systems which do not act as rational, coherent goal maximizers or which have closed-ended goals may be less drawn towards extreme forms of instrumentally convergent behaviors than sometimes assumed (though that hypothesis remains to be corroborated empirically in complex environments). Thus, against the impression of a strict dilemma between risks from misalignment and misuse, it seems (epistemically) possible that some systems may be misaligned but pose a comparatively lower takeover risk.

**3.2 Aligned AGI and misuse catastrophe**

Let us move to the second horn, i.e., the claim that alignment increases the probability of an AGI misuse catastrophe. Friederich (2023) presents a simple conceptual argument for this claim. He claims that "completely aligned AGI, by definition, tries to do what its operators want, whatever that is" (sect. 4). Since aligned AGI, according to this argument, does whatever one wants without constraints, users can employ it for whatever they want, including dominating other humans.

More precisely, we may understand alignment as the correspondence between an AI's goals and the goals of some group $A$ (the "alignment target").[13] The alignment target is the group that the AI system is supposed to be aligned with, which may be designers or users of AI

---

[13] We use the term "goal" here but others may legitimately prefer the terms "values", "preferences" or something similar on the side of the alignment target.

systems, all humans or some other group. Friederich's concern can then be expressed as follows:

Take AGI to be aligned to some alignment target $A$. In this case, every subset of $A$ can use the AGI for their own benefit, thereby frustrating the aims of the members of $A$ complement (the set of elements not in $A$). As aims outside of $A$ are not considered, there are no bounds to how strongly they may get frustrated. We take it that most scenarios in which others' interests are assigned no weight during goal pursuit give rise to misuse. What makes this case different from other cases of unequal access to technology is that AGIs are, by definition, so powerful. For this reason, a group of humans – or an individual – with control over an AGI could use it to suppress the rest of humanity. If one accepts the general argument that misaligned AGI could disempower humanity, then it seems like one should also accept that an alignment target with control over an AGI could use it to disempower the rest of humanity.

Yet, a purely conceptual argument, based on what the concept *alignment* means, is not sufficient to show that efforts at AI alignment increase risks of misuse. Of course, there are strong incentives for any group $A$ to ensure that an AGI's goals aligns with $A$ and for individuals $a \in A$ to use it for their benefit, harming the interests of members of $A$ complement. However, if $A$ is the set of all moral patients (Jaworska and Tannenbaum 2021), then the conceptual argument is undermined. In this case, there would be no moral claims in $A$ complement that will be disregarded by aligning AI with the goals of $A$. Thus, the claim that AGI alignment risks a misuse catastrophe can, on a conceptual level, be countered by taking $A$ to be sufficiently broad.

Crucially, alignment plausibly comes in degrees (Friederich & Dung forthcoming): AI goals can conform more or less faithfully to human goals. Partial alignment reveals new possibilities for navigating catastrophic AI risk. It may be possible to achieve sufficient alignment to avert catastrophic takeover, without making catastrophic misuse likely. Indeed, we advocate for research directions with this goal below. On the other hand, the worst of both worlds is also possible: Some forms of alignment may not reduce the chance of a takeover catastrophe (much) but nevertheless substantially increase catastrophic misuse risk, e.g. because partial alignment may enable single instances of misuse which might already have catastrophic consequences for very powerful AI. While the possibility of partial alignment entails that takeover and misuse risk are not mutually exclusive, this is compatible with a tradeoff in which all other things being equal, increases in AGI alignment decrease the risk of a takeover catastrophe but increase expected misuse risk and vice versa.

There is another way in which alignment admits variation. System *S* is aligned to an alignment target *A* *statically* if *S*'s goals are currently in agreement with *A*'s goals — regardless of whether *A*'s goals will change later on. *S* is aligned to *A* *dynamically* if *S*'s goals are in agreement with *A*'s goals indefinitely, in spite of changes in the latter. Possibly, *S* has the second-order goal to do whatever *A* wants it to do (at any given time point) or its goals are updated reliably to reflect *A*'s goals. In either case, they continue to reflect *A*'s goals (see Green (2024) and Thornley (2024) for this distinction).

Now, given a weaker, merely static form of alignment, catastrophic misuse seems less likely. Whether catastrophic misuse by *A* remains plausible in this case will depend on whether *S*'s and *A*'s goals are in (sufficient) agreement for long enough. If they are not, then *A* cannot reliably control *S* so that intentional multi-step plans for (mis)using *S* cannot be implemented, although – as noted above – partial alignment might suffice for causing catastrophic outcomes. Static alignment may quickly turn into misalignment so that we are back to the risk of an AGI takeover catastrophe. Since dynamic alignment entails that *A*'s and *S*'s goals remain in agreement, it allows for more (and more sophisticated) forms of misuse.

To summarize: We have shown that there is conceptual space for research directions which decrease misalignment risk without making misuse risk more likely. Candidates are research that helps build AI which only follows behavioral heuristics instead of engaging in planning, research that ascertains appropriate constraints for closed-ended goals and the pursuit of a kind of alignment that is sufficiently inclusive to (potentially) avoid misuse risk. Yet, we did not make any claims regarding whether it is empirically plausible that AGI alignment increases the risk of an AGI misuse catastrophe. Hence, there may nevertheless – as a matter of empirical fact – be a tradeoff between alignment and misuse: AI alignment techniques may (typically) reduce AGI takeover risk but increase misuse risk. Section 2 has given an initial reason for this hypothesis: People misusing an unaligned AGI may not be able to implement plans which would cause a misuse catastrophe. In the next section, we look at some current techniques from alignment research to enable an empirically informed assessment of this potential tradeoff.

## 4. The alignment tradeoff in practice

Here's the plan for section 4: First we outline select techniques from empirical alignment research (§4.1) and note present limitations of them (§4.2). Then we analyze scenarios of AI misuse. Based on these analyses, we make some tentative claims regarding our target question: Which concrete AI alignment techniques and research directions, if any, as they are pursued in practice, increase the risk of an AGI misuse catastrophe and by how much? (§4.3). Finally, we draw some broader lessons (§4.4).

We focus here on alignment research which is either based on empirical experiments on AI systems or at least indirectly motivated by empirical data, as this research has come to dominate the field.[14] Currently, much empirical alignment research is performed on large-language models (LLMs), as these constitute the most generally capable class of models which seems most suited for learning about alignment and for concerns about misalignment. This emphasis on LLMs also fits with foreseeable future development of "AI agents" which are centrally built around and directed by an LLM and constitute a major focus of current research (Wang et al. 2024). Moreover, from a current perspective, it also seems plausible that future AGI robots might be directed by some kind of LLM-like system, which is a focus of ongoing research (Ahn et al. 2024; Ichter et al. 2023; Kannan et al. 2024; Kim et al. 2024).

**4.1 Some alignment techniques**

Let's then start with a first class of alignment techniques, namely representation engineering (RE). RE builds on representation reading (or probing) techniques that train an independent classifier to infer the information which is represented within the internal activations of a neural network, without access to its external behavior. RE inserts specific internal representations into the model which are supposed to lead to more desirable behavior (Zou, Phan, et al. 2023). The technique is quite flexible in that by inserting specific representations related to helpfulness, honesty etc., the model can be made more/less helpful, honest, fair or be more/less prone to power-seeking, hallucinations or portraying emotions like anger (ibid.). On the flipside, the model's output can also be made less helpful, more dishonest etc. by inserting corresponding representations.

Following Ji, Qiu, et al. (2024), we call a second class of alignment techniques "learning from feedback". This includes most canonically the fine-tuning technique reinforcement learning from human feedback (RLHF). In RLHF, human preference ratings of evaluators comparing alternative outputs of an AI model are used to shape the model's behavior to get as close as possible to common human preferences. This preference feedback is used to train a separate reward model, the output of which is used to fine-tune the original language model via "classical" RL, i.e. typically proximal policy optimization (Christiano et al. 2017; Ouyang et al. 2022; Schulman et al. 2017; Ziegler et al. 2020). Two recent variations are direct preference optimization (DPO) which is simpler and more efficient in that the AI model is fine-tuned directly based on human preference data without using or training a separate reward model (Rafailov et al. 2023) and constitutional AI: an adaptation of RLHF with the central difference that the feedback used for fine-tuning is generated by a LLM instead of humans (Bai, Kadavath, et al. 2022).

---

[14] We can only briefly review some research directions here. In-depth assessments and surveying the field comprehensively are beyond scope. See Ji, Qiu, et al. (2024) for a recent comprehensive survey of AI alignment research.

As RLHF and constitutional AI are typically conceived and practiced, they aim to make language models helpful, harmless and honest (HHH) (Askell et al. 2021; Bai, Jones, et al. 2022; Bai, Kadavath, et al. 2022). Helpfulness and honesty consist in satisfying the interests or following the instructions of users without deviating from the truth. However, harmlessness requires not producing outputs regarded as dangerous, even if users want them (and they are true).

AI companies employ techniques such as RLHF and constitutional AI to train LLMs to, as best as possible, not produce dangerous information, elaborate lies, racist statements etc., even if the users explicitly ask for them. These techniques for learning from feedback are perhaps the central effort to make models resist attempts of users to produce undesirable outputs via adversarial prompts/attacks (Mowshowitz 2022; Schlarmann and Hein 2023; Shen et al. 2024; Zou, Wang, et al. 2023). When such "jailbreaks", i.e. vulnerabilities to adversarial inputs, were eliminated, the model would be considered aligned, at least in this respect.

**4.2 Limitations of current alignment techniques**

Naturally however, approaches like the "HHH-method" face the problem that the different (here three) principles can conflict (Bai, Kadavath, et al. 2022; Wei et al. 2023). These conflicts are often leveraged in adversarial prompts to circumvent the model's alignment (Millière 2023) to, e.g., misuse it. More generally, given that these adversarial attacks come in so many different varieties due to the vast number of different possible inputs to current LLMs and given that they leverage one of LLMs greatest abilities, namely their ability for in-context learning, there is currently no solution in sight. One might also argue that trends towards bigger models, multimodality and larger context windows will increase the attack surface for such adversarial attacks and thus further magnify this issue (Millière 2023; Shayegani et al. 2024; Wei et al. 2023).

One may hope for advances in techniques for learning from feedback which improve LLMs' resistance to adversarial attacks. However, currently, the ways around such techniques are plentiful. At the same time, the behavioral tendencies instilled by this kind of fine-tuning can be overwritten cheaply via tiny amounts of further fine-tuning, making them appear more like a shallow behavioral gloss than an approach to be trusted for increasingly high stakes (Jain et al. 2024; Ji, Wang, et al. 2024; Qi et al. 2024).

To draw a more general lesson: Despite labs' extensive efforts to avoid misuse of their models, none of them succeed at it, thus making the problem of avoiding misuse seem very hard, and misuse manifesting in higher stakes contexts appear quite plausible. Importantly though, that these alignment techniques fall short of *preventing* misuse does not yet imply that they *facilitate* (catastrophic) misuse, i.e. make it more likely. We have yet to make that case.

**4.3 Alignment techniques and misuse risk**

Here, a starting point is the just-mentioned fact that these techniques are very general and can be used by lots of actors to change the behavioral tendencies of the models, including after thorough alignment (to the goals of all moral patients). Note first that the idea of approaches such as RLHF or constitutional AI increasing misuse risk is not new. Bai, Kadavath, et al. (2022) for example acknowledge that their work has dual-use potential, stating that constitutional AI "lower[s] the barrier to training AI models that behave in ways their creators intend", and makes it "easier to train pernicious systems" (ibid., p. 16). While we focus on AGI and catastrophic risks here, this idea of alignment techniques simplifying the training of worrisome AI systems will be central to our assessment.[15]

A crucial factor is the ease of misusing a given alignment technique. How much alignment raises misuse risk differs between techniques and is centrally based on who can misuse them. This depends on access to the model and thereby importantly on how companies deploy their AI systems. For example, jailbreaking can be done by anybody with access to the model for inference, while RLHF requires at least an API access that enables fine-tuning. Finally, RE requires access to model activations and can therefore – without exfiltrating the model weights (see footnote 15) or them being openly accessible – only be done by designers for now. The form of deployment that is most vulnerable to catastrophic misuse is making model weights generally accessible since it gives all interested people the most comprehensive form of access to a model, thus thwarting most options for implementing effective obstacles to misuse (e.g., it enables reversing safety fine-tuning). Since model access influences the risks posed by different actors, let's discuss those next.

---

[15] We focus on (rather) direct pathways by which alignment research may increase catastrophic misuse risk. For illustration, a scenario which we bracket here is the following: Alignment techniques may be effective, but not sufficiently reliable. This could give a wrong impression of safety which leads to a lack of caution that ends up facilitating misuse of actors either internal or external to leading AI labs (Dung 2024b, sect. 3.1). These scenarios are worth thinking about but depend on different considerations than the cases we focus on.

Clearly, various groups of people could pose misuse risks. To simplify things, we consider two groups (and pathways) for which we later conclude that they vindicate the AGI alignment tradeoff since it is indeed plausible that alignment techniques may increase misuse risks. First, extant aligned AGI could be influenced by *users* to act maleficently. This is analogous to adversarial prompts which are currently used on LLMs like GPT-4 to circumvent its alignment measures so that it, e.g., produces racist statements. Second, *designers* could develop AGI to make it easy (for them) to misuse. They could also modify an originally safe AGI system to align it with their own malevolent goals. Current methods for this are fine-tuning a LLM, overwriting its fine-tuning, or selectively intervening on its internal representations to change its behavior (e.g. Meng et al. 2023) to follow the designer's goals.

We set aside *external actors* which illicitly obtain access to model weights or similarly essential source code. Since this effectively puts them in the position of designers, (largely) the same considerations for the tradeoff between alignment and misuse apply.[16,17]

Let us focus on misuse by users first. It is plausible if not overwhelmingly likely that among all the users, some will have nefarious purposes (see e.g. Marchal et al. (2024) for an overview of recent misuse tactics). The question is whether labs can prevent users from misusing AIs towards such ends. Much depends on the extent and sophistication of labs' safety efforts and the resources and skills of the user. We mentioned above that misuse of models by users is common and that, presently, there is no compelling reason to expect that it will be possible to effectively prevent misuse in the future.

Are nefarious users likely to cause catastrophic consequences, though? Consider the following: Specific AI systems can currently be used to predict new highly toxic chemicals (Urbina et al. 2022). Preliminary evidence suggests that current LLMs may facilitate access to hazardous pathogens (Soice et al. 2023) though two subsequent red-teaming studies did not find a significant difference in the viability of plans to execute a biological attack between groups with only internet access and those with additional access to an LLM (Mouton et al. 2024; Patwardhan et al. 2024). It does, however, remain plausible that LLMs will rather soon become comparatively useful at such tasks and more capable in this domain when they, e.g., incorporate more specialized knowledge, for example about virology; see Anthropic (2025a), OpenAI (2025), and especially Anthropic (2025b) for recent evidence of such comparative usefulness.

---

[16] Cyberattacks to exfiltrate model weights nonetheless constitute a very important factor contributing to misuse risk. Incentives for such attacks are high and rising as the costs to train leading AI models are already at $100 million (Maslej et al. 2024). Cyberattacks are common in many profitable sectors and even organizations that invest heavily in security such as the NSA, CIA, Google and Microsoft had severe breaches before (Nevo et al. 2024). Again, with stolen model weights, malignant actors are effectively in the position of AI designers and can, using alignment techniques, adapt very powerful systems like AGI to serve their preferences with corresponding consequences.

[17] Similarly, if users can fine-tune the model through an API and particularly if the model is leaked or open-sourced, then users are in a similar position to designers. Using alignment techniques, they can adapt the AI's behavior and goals in accord with their preferences which constitutes a misuse risk.

However, current LLMs may nevertheless be insufficient to enable widespread access to biological weapons because creating them requires laboratory know-how and access to lab-equipment that most actors which would benefit from LLM access cannot obtain. It still follows by definition that an AGI system would have capacities which suffice to cause massive damage via biological weapons, given that groups of very capable humans are able to do so. In this scenario, AI companies will likely try lots of techniques to prevent their models from performing such hazardous actions (following user prompts), for instance when robots are capable of directly synthesizing dangerous chemicals, or from facilitating such actions by spreading hazardous information. If the safety of such AI systems to, e.g., adversarial prompts is still porous, catastrophic misuse by some ill-motivated actor eventually seems likely.

How could alignment techniques increase such risks from users? By making systems more helpful and honest, techniques for learning from feedback make it more likely that LLMs would spread hazardous information. While harmlessness acts as a counterweight, a system which is not sufficiently fine-tuned for honesty or helpfulness is arguably much less dangerous in the first place, even if no special harmlessness training is conducted. To the extent that representation engineering can be employed by users, the same concern arises: This alignment technique makes AI systems follow the users' instructions and wishes more precisely. Hence, if users have dangerous aims, this technique is liable to make AI systems more dangerous, by increasing misuse risk.

Now, assume that extant techniques to circumvent safety measures – such as adversarial prompts, overwriting previous alignment via further fine-tuning and similar strategies – are not effective in the AGI case anymore (for instance, if and because AGI is more agentic – see below) or that companies have found other methods to defend against these approaches. One may still worry that users can find holes in their defenses or novel strategies for misuse. For example, not only are LLM-controlled robots vulnerable to extant strategies for jailbreaking (Ravichandran et al. 2025; Robey et al. 2024.), they might also have new vulnerabilities. Here, we can only say that, in general, since the users include state actors aiming for misuse (Microsoft 2024; Nimmo 2024), their abilities should not be underestimated.

Let's then discuss misuse by designers. It is evident that this remaining scenario is also possible. Indeed, it is the most straightforward scenario. Most importantly, designers can apply alignment techniques to change the model at will and according to their preferences. This includes cases where the AGI is initially aligned since both RE and techniques for learning from feedback can be used to overwrite its goals and instill undesirable goals in it. By profession, AI designers are very skillful at adapting the behavior of AI systems.[18] If AGI is aligned with its designers, then its designers gain massive power and can use it to, e.g., subjugate all other humans.

The misuse risk here may be limited by the relevant group of designers being sufficiently moral or sufficiently tightly controlled to not misuse AGI. At least in current practice, designers do not align AI systems to do whatever they tell it to do as e.g. GPT-4 resists outputting answers considered harmful even if its designers ask it to. However, this still appears like a very unstable and dangerous situation as it depends on the good will or sufficiently tight control of designers and remains vulnerable to the same issues detailed above for users.

In addition, the control of designers is surely a double-edged sword. We must also consider that designers may be controlled e.g. by malevolent governments, corporations or dictators and thus drop the often-implicit assumption that models are trained and controlled by relatively harmless (groups of) people. Clearly, such malevolent actors may coerce the designers to adapt the AGI system via alignment techniques to facilitate misuse or directly towards malevolent ends and are plausibly very motivated to seize power in such a way. In all cases detailed here, directly applying alignment techniques without interference of an external actor plays a central role in the manifestation of a catastrophic risk.[19]

While a higher degree of agency of AI systems (Dung 2024c) increases various risks (Anwar et al. 2024; A. Chan et al. 2023), it may decrease the probability of misuse, at least when considering AGI. Assume that the AGI pursues some fixed set of goals coherently, autonomously, and over long timescales. If it has the goal to resist misuse, then – in virtue of its superhuman capacities – we should expect that it widely succeeds at preventing being misused.[20]

---

[18] A further important scenario which we do not discuss is the following: Even if the AGI is and remains aligned, one could use the same techniques (perhaps in combination with stealing the model weights of the AGI) to align a different AGI with malicious goals. If this second AGI system causes a catastrophe, this catastrophe has been facilitated by alignment research.

[19] While we focus for simplicity on just a single AGI system here, the scenarios discussed could also manifest as cumulative as opposed to decisive catastrophic risk (Kasirzadeh 2024) where many AIs get increasingly capable and are increasingly often and severely misused, ultimately leading to catastrophic consequences.

[20] That being said, even if an AGI is generally more capable than humans, having the goal to resist misuse does not rule out vulnerabilities to very specific attacks via which its behavior can be controlled. For example, perhaps

This does not afford total protection against catastrophic misuse, however, because alignment including intentional misuse resistance might only be partially achieved. During alignment training, researchers may not place enough emphasis on the varieties of potential human misuse which could, e.g., circumvent or overwrite previous alignment using the same techniques so that the system may afterwards not try to resist misuse. As noted above, at present, users commonly find ways to make AIs behave contrary to what, say, RLHF was supposed to achieve. In particular, misuse resistance may be harder to achieve for AI agents as they provide a bigger attack surface than LLMs, making them currently particularly vulnerable (Andriushchenko et al. 2025; Greshake 2023; Zhang 2024). Putting both concerns together, the worry is an outcome where alignment techniques suffice to make an AGI system controllable – it follows human instructions and can systematically accomplish their goals, making misuse possible in the first place – but has weaknesses in not robustly resisting misuse attempts, enabling some users to direct the system at undesirable goals instead.[21]

Moreover, an agentic AGI may not try to resist misuse. Indeed, humans may want to design AGI such that it is as useful as possible for achieving their goals. Who wants an AI assistant who intentionally resists doing what one asks of them? Aside from mundane (economic) utility considerations, some AI safety efforts also push in a similar direction. For example, corrigibility, a common goal in AI safety, means that the system can be "corrected" and hence also repurposed from the outside (Lo et al. 2019; Russell 2019; Soares et al. 2015). In this case, as the AI's goals can be corrected with whatever the new alignment target wants, the AI can be enlisted to help with arbitrary outcomes, including catastrophic ones.

If the AGI is not agentic, we can rule out intentional resistance by the model itself so that the worry is directly that alignment techniques such as RE or RLHF (in combination with jailbreaks) can be used to control the system to misuse it. So, the previous considerations apply. Consequently, if an AGI is not particularly agentic, one potential obstacle for misuse is removed such that misuse risks are more pressing.[22] With respect to less capable systems, however, it seems plausible that AI agents remain vulnerable to misuse attempts.

Let us, finally, briefly mention an additional approach to AI safety: robustness. Generally, robustness concerns the ability of AI systems to maintain reliable performance in

---

it is possible to insert a "backdoor" into the AGI while it is trained such that the AGI is not aware of it and that it can later be exploited to control the AGI's behavior (Hubinger et al. 2024).

[21] While AGI misuse may often be maximally severe, when considering less capable systems, agency increases the severity of potential misuse incidents as AI agents can act autonomously towards dangerous objectives given to them.

[22] This contrast suggests that agency as a dimension distinguishing different future AIs plausibly encompasses an independent tradeoff between misalignment and misuse risk: lower agency reduces risks from misalignment (Dung 2024c) while it raises misuse risk.

spite of various challenges such as inputs that are outside of the training data distribution, including adversarial inputs (Carlini et al. 2019; Hendrycks et al. 2021; Ren et al. 2020). One approach in this research direction is to ensure that the goals of AI systems remain stable under distribution shift (Lubana et al. 2023; Ziegler et al. 2022). This research direction could hence, if sufficiently successful, exclude the possibility that models are misused via adversarial prompts. Moreover, since robustness research does not aim at directly changing the kinds of behaviors or goals of AI systems, but only at making their behavior more stable, there may be no correspondingly direct causal paths from such research to a misuse catastrophe.[23]

## 4.4 Upshot

Let us summarize and return to a bird's eye view. The basic argument why alignment techniques may contribute to a misuse catastrophe is that without any previous alignment, catastrophic misuse seems extremely hard, perhaps in many cases practically impossible. This is because alignment techniques are essential for making AI behavior predictable and useful.

A second important general point is that numerous alignment techniques in practice seem to enable fine-grained control of AI systems' behavior or goals. This has the important consequence that it seems quite plausible for alignment techniques to facilitate catastrophic misuse risk.[24] This brings us back to the two forms of alignment from section 3.2. First, one may wonder whether current alignment techniques are suited to achieve only static or also dynamic alignment. Focusing on learning from feedback as an example, one may argue on the one hand, that instruction fine-tuning in particular aims to get the model to follow arbitrary instructions, changing AI behavior when the instructions change, and hence fits better with a dynamic understanding of alignment. On the other hand, one may describe the application of learning from feedback methods usually as getting the AI's goals and behavior to conform better with a particular, static set of preferences regarding model behavior (as generated by humans or an AI). We make no claims here about which of the two factors is decisive as that requires empirical research. As static alignment is more prone to risks from misalignment while dynamic alignment incurs a higher misuse risk, we, however, want to flag this uncertainty as an important issue for further research.

---

[23] Plausibly though, perfectly robust systems cannot be redirected so that it may *indirectly* raise risks from misalignment if the goals or behaviors are initially chosen badly.

[24] It should also be noted that a conceptual account of alignment that is quite broad (though usually restricted to humans) is commonly championed by leading AI companies. At the risk of psychologizing, this is at least a worrying pathway for safety-washing since the abstract account may be misleading regarding what alignment is about in practice.

Second, on the conceptual level it seemed that, for an AGI aligned to a group *A*, misuse risk could only be increased for moral patients outside of *A* (and may even be excluded for a sufficiently inclusive kind of alignment). However, in practice, many different agents may determine or overwrite the behavioral tendencies of the A(G)I system using alignment techniques to make them conform to their goals. Thus, all moral patients – whether they are in the initial alignment target *A* or not – may in practice face misuse risks from various agents. Hence, the just-sketched understanding of alignment techniques and misuse scenarios in practice suggests higher misuse risk than the abstract notion of alignment suggested above.

An important limitation of our argument is that we focus on few specific techniques and research directions in AI alignment research here. While an analysis of how these may be applied gives a first impression of how likely they are to lead to catastrophic misuse, we naturally have excluded other research directions. Particularly for such a young and quickly developing field, one cannot predict reliably which practices will be common in a few years. Further research is needed here.

In total, it appears likely that some influential alignment techniques — to the extent that they are effective at their intended uses — indeed increase the chances of an AGI misuse catastrophe, although the matter is necessarily speculative. From an empirical perspective, the AGI alignment tradeoff is plausible, possibly even more so than from a conceptual perspective alone.

## 5. The social context of AI alignment and misuse

The risks of AI cannot be evaluated from an analysis of the technical capacities of AI systems in isolation — attention must be paid to the social context in which AI systems are developed and deployed (Bolte et al. 2022; Johnson and Verdicchio 2024; Weidinger et al. 2023).[25] Systemic, societal factors forge the risk landscape, e.g. by shaping which technologies are developed (first), which safeguards and safety methods are researched, developed, and enforced, how or if AI systems are deployed and monitored, and especially important here, how they are (mis)used.[26]

A common categorization of risks into misuse and accident risks (where accident risks include misalignment) neglects structural risks. Examples of the latter are privacy erosion as AI increases the value of data, the drive towards autonomous weapons as militaries mutually fear that their opponents develop them, or widespread AI-driven labor displacement. These illustrate that some risks do not fit squarely on either side of the accident-misuse-dichotomy or as just mentioned, shape these other risks (LaCroix & Mohseni 2022; Zwetsloot & Dafoe 2019). In addition, since both misuse and misalignment risks are often analyzed as though they are abstract, isolated phenomena, particularly in the case of AGI, it is plausible that the neglect of contextual factors is responsible for the near universal lack of appreciation that AI alignment may facilitate misuse risks — the central issue at stake here.[27] Hence, this section deepens the emphasis on social factors surrounding both risks.

---

[25] We are thankful to an anonymous reviewer for encouraging us to go into greater detail in analysing social factors relevant to AI risk.
[26] It is worth adding that human goals as the alignment target are an inherently social and dynamic phenomenon.

We first address the question whether social factors may improve catastrophic takeover risk without increasing catastrophic misuse risk, and vice versa. Call these "uniform improvements". Second, we briefly discuss broader social factors which substantially shape or constitute the AGI risk landscape we face.

We stress that it is hard to predict the social context in which AGI will be developed and deployed. A central consideration is the quality and focus of AI governance, including the form or lack of international coordination which we cannot anticipate in advance. Even central features of the future governance landscape, for instance whether AGI will be developed by private firms or eventually by governments, are open. This complicates the prospects of a rigorous socio-technical analysis of AGI risk. Nevertheless, we think that some important, albeit preliminary, insights can be reached.

First, for many social factors, it is plausible that they reduce the probability of a takeover and misuse catastrophe. It is commonly noted that AGI development driven by strong competitive race dynamics, between firms or states, incentivizes reckless and unsafe development (Armstrong et al. 2016; Hendrycks et al. 2023; Katzke and Futerman 2024). Race dynamics increase the chances for all kinds of risks and reducing such dynamics should improve risk management across the board. Hence, social interventions that reduce race dynamics may be uniform improvements. By intentionally stopping or decelerating particularly dangerous kinds of AI development (Cohen et al. 2024) one could for example provide social systems, including regulatory bodies, more time to intentionally steer AI development and adapt to downstream consequences thereof (Bernardi et al. 2024).[28] Other governance proposals, such

---

[27] A similar point may hold for the fact that this tradeoff between misalignment and misuse risks has not been itself recognized (in the few places that it has been discussed) as a contingent (complex) social phenomenon but has instead rather been treated as an abstract conceptual fact, against which we argued in section 3.2.

[28] The need for sufficiently fast and adaptive institutions to govern increasingly capable AI is a central recurring issue here. A particular challenge is that governance proposals risk being either based on too little evidence making them ineffective or counterproductive or alternatively, being too late to mitigate risks appropriately when waiting for stronger evidence (Bengio et al. 2025).

22
This is a preprint. Please cite the published version when available.as mandatory risk assessments before AI deployment by third parties with comprehensive access (Casper et al. 2024; Raji et al. 2022; Shevlane et al. 2023), clarifying liability for AI harms, compulsory reporting of safety cases[29], and know-your- customer requirements (Egan & Heim 2023), may be further uniform improvements. Defense in depth as an increasingly popular approach combines many independent risk mitigation measures like the ones suggested here (and many more) to drive down overall risks despite no single measure being able to provide or assure safety (Bengio et al. 2025; Hendrycks 2025). While opinions vary on how strongly each measure decreases takeover or misuse risk, thinking that one will be very effective at reducing one of these risks does not provide a (clear) reason to think it will increase the other. Since social and governance interventions plausibly decrease either of these risks without increasing the other, there is a defeasible but substantive reason to put more weight on such interventions over narrow technical ones.

Second, some social interventions may trade off reductions in one risk for increases in the other. Most simply, if governments decide to massively increase funding for, say, alignment research, this intervention faces the alignment-misuse tradeoff concerning current alignment techniques for which we just argued. A more substantive question concerns the release of (advanced) AI (Seger et al. 2023; Solaiman 2023), e.g. by making the weights of powerful models openly accessible. On the one hand, publishing the weights of powerful models makes it harder to ensure that they will be deployed in an aligned manner. It also gives more actors a chance to misuse them. On the other hand, for now, widely distributed models facilitate alignment and safety research, particularly by under-resourced and independent actors while concentrating powerful AI in the hands of few actors risks massive power concentration; a crucial factor for catastrophic misuse risk. So, while uncertainty abounds now on how releasing powerful open-weight models would affect catastrophic takeover and misuse risk, it is likely that in some realistic situations, both risks trade off against each other. In total, though, social factors may frequently be uniform improvements (or at least affect both risks in the same direction) instead of trading off misuse risk for misalignment risk or vice versa.

Let us briefly cover some more general factors regarding the social context of AGI risk. First, incentives play a quite general role beyond the foregoing discussion of competitive race dynamics for the development of increasingly advanced AI. To begin, financial incentives for researchers supplied by industry and the corresponding potential influence (even absent manipulative intent) have grown substantially and are particularly pronounced in (various fields of) AI (Abdalla et al. 2023; Abdalla & Abdalla 2021). We deem it important, particularly for instituting impartial governance, that enough (well-resourced academic) researchers remain (sufficiently) independent of such incentives from big tech corporations. Further incentives concern AI companies which have an interest to develop AI systems in rather specific ways. With regards to the risks we focus on here, AI companies are incentivized to e.g. align AIs to users to maximize usefulness for them and consequently the amount they are willing to pay, even if that compromises measures against misuse (or misalignment).[30] Similarly, and most obviously, companies, developers, and nations are incentivized to train AI systems that privilege their particular interests over those of others. At a minimum, we reckon that companies ought to transparently document sufficient information about their training to assess how they balance company/developer interests with those of users and external parties to enable public scrutiny.

---

[29] Safety cases are structured arguments to the conclusion that a system is sufficiently safe (regarding certain risks) in a given operational context (Buhl et al. 2024; Clymer et al. 2024).

Second, on a meta-level, there is a question about the (comparative) leverage of social interventions on misuse and misalignment risk. Here, we contend that misuse risk is more strongly influenced by social context than instantaneous takeover risk. The reason is simple: Once a sufficiently powerful model with the goal to disempower humanity is released, our social arrangement will by assumption not be able to constrain it and thus have little influence on the outcome. In contrast, misuse is inherently a sociotechnical challenge (Anwar et al. 2024). Whether misuse occurs depends on many social factors, e.g. the motivations of various actors, how widely AGI access is distributed, which other resources relevant actors have at their disposal and how their actions are monitored, constrained, and so forth.

However, the social context is also crucial to risks from misaligned AGI if the outcome in question does not unfold suddenly but gradually. A gradually accumulating existential catastrophe has been hypothesized to follow either a series of AI-induced disruptions which lower the resilience of different societal systems until a cascading failure of these systems causes irrecoverable collapse (Kasirzadeh 2024) or as gradual takeover because competitive incentives and coordination failures drive humans to give up gradually more control to AI, until societal systems are decoupled from human feedback and control (Kulveit et al. 2025). The first risk scenario closely depends on the overall resilience of different societal systems, while the latter requires remedying potential coordination problems to ensure that humans do not hand over undue amounts of control to AI.

---

[30] This does not mean that there are no incentives to mitigate these risks. For instance, misuse can create PR damage and encourage strict regulation. Additionally, companies may be held liable for damages.

Finally, one may wonder about societal mitigation efforts. How many resources will be put to misuse mitigation and alignment research in AI labs, academic institutions, NGOs and governments; both in total and relative to investments advancing AI capabilities?[31] Of course, this is an open question but, plausibly, it will depend centrally on a more general appreciation and understanding of risks from AI including their social context. Here, it is also noteworthy that physical environments are often more strongly regulated than virtual ones, suggesting that a proliferation of AI-controlled robots might cause stronger social responses to AI risks. To summarize, social factors overall provide a promising lever to address AGI misuse and misalignment risks as they rarely seem to trade decreases in one for increases in the other. By shaping the incentives for the involved players, e.g. by mitigating race dynamics, mandating risk assessments, or clarifying liability for harms by AI, governments can influence the behavior of relevant actors to better align with the common good.

## 6. Conclusion

We started by showing that there may be a tradeoff between risks due to misalignment and catastrophic misuse as both horns in support of this conclusion could not be rebutted. However, we also suggested directions in which – at least in the abstract – misalignment and misuse may be reduced without trading off against each other. Therefore, aligned AGI does not conceptually *imply* a high misuse risk. Hence, whether alignment efforts will increase misuse risk is a contingent empirical question. We argued that some current alignment techniques and foreseeable improvements thereof plausibly increase risks of catastrophic misuse. In addition, while the AI misuse risks seemed constrained on the conceptual level to specific groups and notions of alignment, considering alignment techniques and plausible threat models suggest that no such constraints are likely in practice. While we argued that the risks we discuss are even more pronounced when encountering AI-controlled robots, an increasing presence of robots may also have various second-order consequences, including stronger social responses to AI risks since robots increase their saliency.

Of course, that some alignment efforts may increase catastrophic misuse risk is not by itself a sufficient reason to give up on the goal of AI alignment. Even if it turned out that alignment techniques inevitably increase misuse risk, they may still be worth using in practice, depending on the magnitude of the increases and reductions in risks from misuse and misalignment.

---

[31] Particularly so as only between less than 1% and 2% of AI publications are on AI safety (Emerging Technology Observatory 2025; Toner & Acharya 2022).

In addition, a remaining question is whether there are different approaches in AI safety or AI alignment research that reduce or avoid this tradeoff. While this is an open empirical question, we think that currently, robustness research seems like a good candidate for reducing the tradeoff. A further promising direction may be methods from AI control research as, e.g., suggested by Greenblatt, Shlegeris, et al. (2024), although implementing them while avoiding other downsides is non-trivial. To prevent misuse in particular, these methods could (for example) make use of more specific information about who uses which AI systems in which ways (Chan et al. 2024). Misuse risks should also inspire a more nuanced evaluation of some proposed convergent instrumental goals. Goal stability, the hypothesized tendency of systems to preserve their own goals, has been identified as a risk factor for an AGI takeover catastrophe, but may make some forms of misuse harder. So, research should explore whether these hypothesized tendencies can be used to reduce misuse risk, without disproportionately increasing takeover risk.

On a broader structural level, our argument provides motivation for a research program which may be called *worst-case AI safety*. This program would search for safety and alignment strategies which reduce the risk of a takeover catastrophe without designing techniques which enable fine-grained behavioral control of AI systems, since such techniques may increase the risk of a misuse catastrophe. In addition, we think that safety evaluations of AI systems (e.g. before deployment) should routinely include risks of misuse by designers and by adversaries which got access to the system (e.g., by stealing model weights), not just users. Further, technical research papers on AI alignment should standardly contain a short assessment of dual-use risks, detailing whether the developed alignment technique increases the risk that irresponsible or malicious actors misuse AI systems in dangerous ways. Finally, since much of the risk for misuse depends on the social context including who uses alignment techniques to adapt AI systems, ensuring good governance and transparency about how AI systems are adapted are essential going forward.

**Acknowledgements**

For very helpful comments on earlier versions of this paper we are grateful to Eleonora Catena, Michael Eigner, Simon Friederich, Peter Kuhn, Isidor Regenfuß, Ian Robertson, Edward Young, and anonymous reviewers.

284